\documentclass[sn-aps]{sn-jnl}

\jyear{2021}%

\theoremstyle{thmstyleone}%
\theoremstyle{thmstyletwo}%

\theoremstyle{thmstylethree}%

\begin{document}

\title[A Simple Experimental Setup for Simultaneous Superfluid-response and Heat-capacity
Measurements for Helium in Confined Geometries]{A Simple Experimental Setup for Simultaneous Superfluid-response and Heat-capacity
Measurements for Helium in Confined Geometries}


\author*[1,2]{\fnm{Jun} \sur{Usami}}\email{jusami@crc.u-tokyo.ac.jp}

\author[2]{\fnm{Ryo} \sur{Toda}}

\author[2]{\fnm{Sachiko} \sur{Nakamura}}

\author[1]{\fnm{Tomohiro} \sur{Matsui}}

\author*[1,2]{\fnm{Hiroshi} \sur{Fukuyama}}\email{hiroshi@phys.s.u-tokyo.ac.jp}

\affil[1]{\orgdiv{Department of Physics}, \orgname{The University of Tokyo}, \orgaddress{\street{7-3-1}, \city{Hongo, Bunkyo-ku, Tokyo}, \postcode{113-0033}, \country{Japan}}}

\affil[2]{\orgdiv{Cryogenic Research Center}, \orgname{The University of Tokyo}, \orgaddress{\street{2-11-16}, \city{Yayoi, Bunkyo-ku, Tokyo}, \postcode{113-0032},\country{Japan}}}



\abstract{Torsional oscillator (TO) is an experimental technique which is widely used to investigate superfluid responses in helium systems confined in porous materials or adsorbed on substrates.
In these systems, heat capacity (HC) is also an important quantity to study the local thermodynamic properties.
We have developed a simple method to incorporate the capability of HC measurement into an existing TO without modifying the TO itself. 
By inserting a rigid thermal isolation support made of alumina and a weak thermal link made of fine copper wires between a standard TO and the mixing chamber of a dilution refrigerator in parallel, we were able to carry out simultaneous TO and HC measurements on exactly the same helium sample, i.e., four atomic layers of $^4$He adsorbed on graphite, with good accuracies down to 30\,mK.
The data reproduced very well the previous workers' results obtained independently using setups optimized for individual measurements.  
This method is conveniently applicable to a variety of experiments where careful comparisons between results of TO and HC measurements are crucial.
We describe how to design the thermal isolation support and the weak thermal link to manage conflicting requirements in the two techniques.}




\maketitle

\section{Introduction}
\label{intro}

Torsional oscillator (TO) is a powerful tool to investigate superfluid responses in helium systems, particularly those in confined geometries or reduced dimensions.
TO measurements can determine the superfluid density, the transition temperature at which the system acquires the macroscopic quantum coherence over the whole sample, a possible dissipation associated with the transition.
Heat capacity (HC) is also an important quantity to be measured, providing us with local thermodynamic information of the system.
In general, HC measurements are sensitive to phase transitions such as Bose-Einstein condensation, liquefaction, and solidification.
By combining these two complementary measurements, important insights have been yielded into the behavior of superfluid $^3$He in aerogel~\cite{Porto1995,Choi2004}, superfluid $^4$He confined in nanoporous materials~\cite{Yamamoto2004,Yamamoto2008}, and superfluid $^4$He films adsorbed on nanoporous substrates~\cite{Berthold1977,Finotello1988,Toda2007}.

Usually, TO and HC measurements are carried out using different experimental setups because they have conflicting technical requirements, e.g., strict mechanical rigidity and switchable or proper thermal conductance.
Sometimes this causes a problem especially when one has to compare the results in great detail.
This is because the porous materials and substrates used in the TO and HC measurements are not exactly the same and may have slightly different characteristics due to different treatment or rot.
One serious example is the second layer of $^4$He adsorbed on a surface of exfoliated graphite, like Grafoil, where possible coexistence of superfluidity and spatial order is anticipated from the previous TO~\cite{Crowell1996, Shibayama2009, Nyeki2017, Choi2021} and HC experiments~\cite{Nakamura2016}.
It is known that 10--15$\%$ of a Grafoil surface is heterogeneous and that helium atoms trapped there behave quite differently from those on an otherwise atomically flat surface~\cite{Sato2012}. 
This results in large uncertainties in the density scales between the different groups.
Thus, it is not easy to firmly assign which phase shows superfluidity in the phase diagram determined by the HC measurement~\cite{Nakamura2016}.
Simultaneous TO and HC measurements on exactly the same sample will solve this problem and be useful for other superfluid helium systems.

So far, there are only a few experiments in which simultaneous TO and HC measurements have successfully been done~\cite{Murphy1990,Csathy1998,He2002}.
In this article, we describe details of the design and test results of a simple method we developed to incorporate the capability of HC measurement into an existing TO without modifying the TO setup to make the simultaneous measurements possible easily in a wide temperature range from 1.5\,K down to $T = 30$\,mK or even lower.

\section{Design of experimental apparatus}
\label{method}

\begin{figure}[b]
\centering
\includegraphics[width=0.75\linewidth]{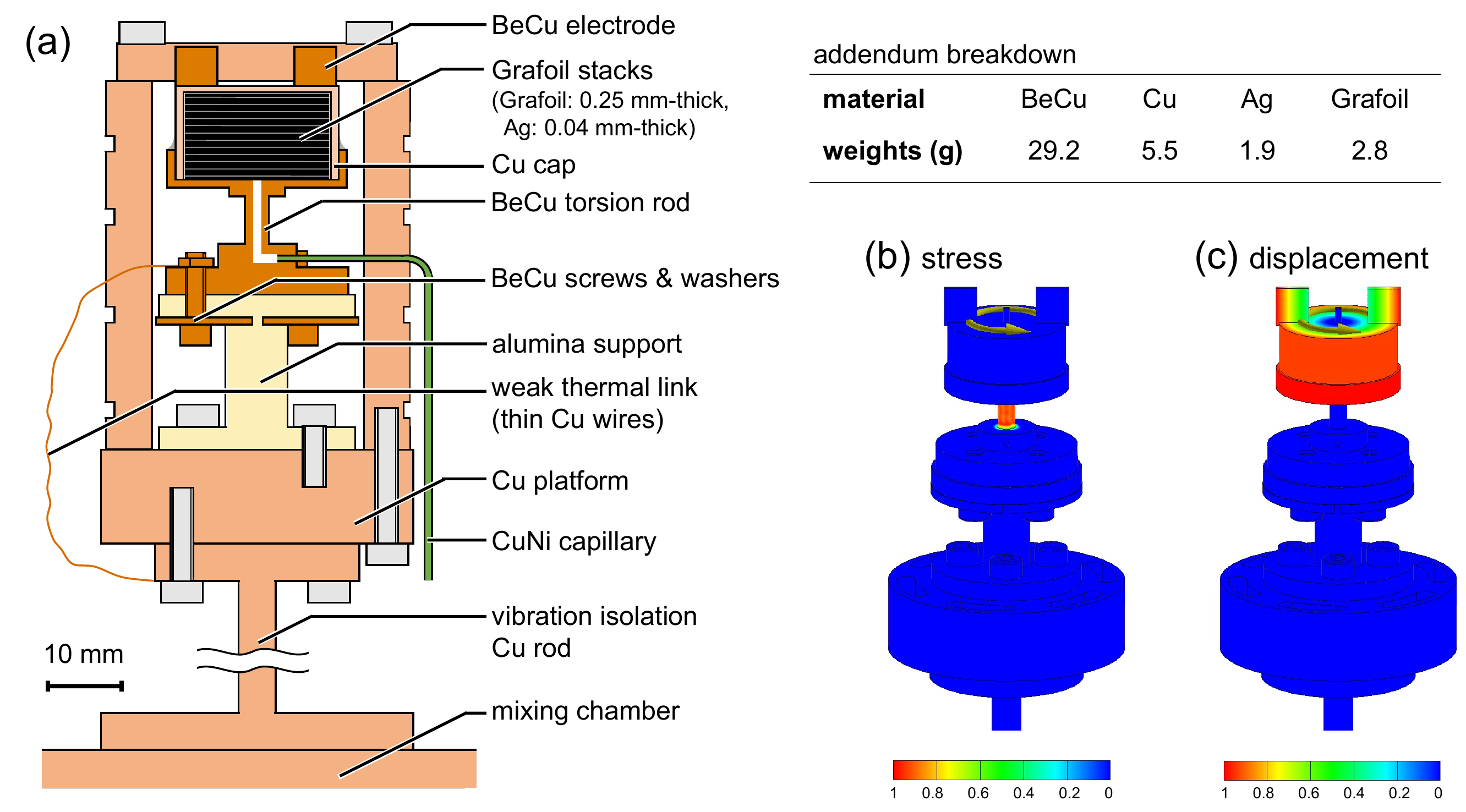}
 \caption{(a) Experimental setup for simultaneous measurements of torsional oscillator and heat capacity. Finite-element method simulations of (b) stress and (c) displacement.}
\label{fig-setup}
\end{figure}

Fig.~\ref{fig-setup}(a) shows a cross-sectional view of the experimental apparatus we developed for simultaneous TO and HC measurements of a few atomic layers of helium adsorbed on Grafoil.
The design of the TO part is essentially the same as that used in our previous experiment~\cite{Kubota2014a} except for the following two points: 
(i) The materials of the torsion rod and the exfoliated graphite substrate are changed from coin silver and ZYX to BeCu-25 and Grafoil, respectively.
(ii) The Grafoil/Ag/Grafoil stacks are now attached to the inner wall of a thin copper cap, not the main body of the TO, and the cap is soft soldered to the BeCu main body rather than glued.
The adsorption surface area of the Grafoil substrate with 21 ventilation holes of 1\,mm diameter was determined as $45.7\pm0.3$\,m$^2$ from an N$_2$ adsorption isotherm measurement at 77\,K.
At low temperatures, the resonant frequency ($f$) and the quality factor ($Q$) were 1392\,Hz and $5\times 10^5$, respectively.

To make HC measurement possible with the TO of this size, we need to decouple it thermally from its surroundings.
At the same time, the TO has to be supported as rigidly as possible for stable TO measurements with high $Q$.
To meet these conflicting requirements, we chose to connect the TO and a massive copper platform for vibration isolation with a rigid support made of high-purity alumina using BeCu screws on the TO side (see Fig.~\ref{fig-setup}(a)).
We carefully designed dimensions of the alumina support so that its maximum applied stress is less than 6\% of that on the torsion rod and that its maximum displacement is less than 2\% of that of the TO cell using a commercial simulator (Inventor, Autodesk, Inc.) as shown in Figs.~\ref{fig-setup}(b) and (c).
This design satisfies requirements not to exceed the elastic limit of alumina and to set the parasitic resonant frequency of the alumina far from the TO resonance frequency.
We need to measure HC precisely in the temperature range of $0.2 \leq T \leq 1.5$\,K for our specific purpose.
Therefore, the support should have a thermal conductance at least three times lower than that of the torsion rod at 1\,K.
From all these conditions, we finally determined the dimensions of the alumina ($ 99.8\%$ purity; Sumitomo Chemical Co., Ltd.) support as 8\,mm in diameter and 14\,mm in length in the central part.

To reduce the addendum HC, we fixed the TO cell to the alumina flange with six screws and washers made of hardened BeCu, not of stainless steel.
Each BeCu washer consists of two semicircular rings not to break the alumina flange by differential thermal contractions of alumina and BeCu.
For HC and related thermal conductance measurements, we used two types of resistance thermometers, Cernox (CX-1010- SD-HT, Lake Shore Cryotronics, Inc.) at $T > 0.1$\,K and RuO$_2$ (RCL471J50, Alps Electric Co., Ltd.) at $T<0.1$\,K.
The Cernox thermometer and a thin-film metal resistor (Susumu Co., Ltd.) heater were varnished on the copper cap of the TO.
The electrical resistance (1\,k$\mathrm{\Omega}$) of this heater changes only 0.3\% from room temperature to 30\,mK.
The RuO$_2$ thermometer made by wrapping a 470\,$\mathrm{\Omega}$ RuO$_2$ chip with a silver foil of 0.04\,mm thick was fixed on the bottom plate of the TO.

For our research purpose, we should extend the TO measurement at least below 40\,mK.
The thermal conductance of alumina $K_{\rm{al}}$ decreases rapidly as $K_{\rm{al}} \propto T^{2.7}$ with decreasing $T$~\cite{Locatelli1976}.
Thus, we added a weak thermal link consisting of three fine copper wires (0.1\,mm in diameter and 100\,mm long each) in between the TO cell and the copper platform, i.e., in parallel with the alumina support, as shown in Fig.~\ref{fig-setup}(a).
As the thermal conductance of this copper weak link $K_{\rm{Cu}}$ should vary as $K_{\rm{Cu}} \propto T$, it will dominate the total thermal conductance, $K = K_{\rm{al}} + K_{\rm{Cu}}$, below about 0.5\,K, while $K_{\rm{al}}$ does at higher temperatures.

\section{Thermal conductance measurements}
\label{thermal_conductance}

\begin{figure}[t]
\centering
 \includegraphics[width=0.7\linewidth]{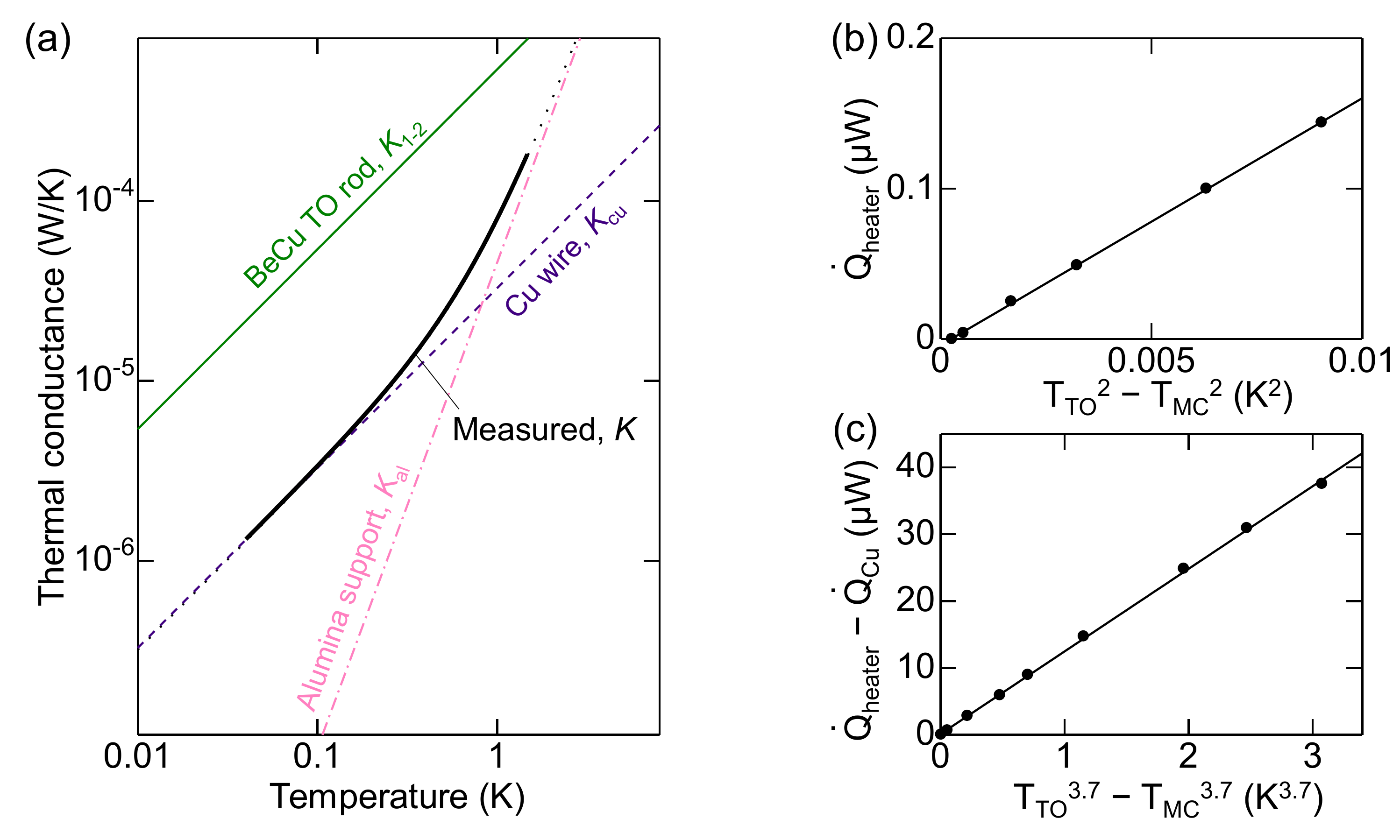}
\caption{(a) Measured thermal conductance between the TO and the mixing chamber ($K$; solid line), and estimated thermal conductances of the copper-wire weak link ($K_{\mathrm{Cu}}$; dashed line), alumina support ($K_{\mathrm{al}}$; dash-dot line) and BeCu TO rod ($K_{\textrm{1-2}}$; upper solid line).
$K_{\mathrm{Cu}}$ and $K_{\mathrm{al}}$ are derived from the data fittings in the low-$T$ (b) and high-$T$ regions (c), respectively. For further details, see the main text.}
\label{fig-K}
\end{figure}

In Fig.~\ref{fig-K}(a), the thick solid (black), dashed (purple), and dash-dot (magenta) lines represent $K$, $K_{\rm{Cu}}$, and $K_{\rm{al}}$, respectively.
They are determined by monitoring the temperature of the TO cell ($T_{\mathrm{TO}}$) and that of the mixing chamber of the dilution refrigerator ($T_{\mathrm{MC}}$) under a constant heat flow ($\dot{Q}$) generated by the heater at steady states through the following procedures.
By fitting the data below 0.1\,K, where $K$ is thoroughly dominated by $K_{\rm{Cu}}$, to $K_{\rm{Cu}}/T=2\dot{Q}/(T_{\mathrm{TO}}^2-T_{\mathrm{MC}}^2)$, we first determined $K_{\rm{Cu}}$ (see Fig.~\ref{fig-K}(b)).  
From this $K_{\rm{Cu}}(T)$, we could estimate the heat flow through the copper wires ($\dot{Q}_{\mathrm{Cu}}$) at any temperatures as long as the $K_{\rm{Cu}} \propto T$ relation is held.
Then, using the heat flow through the alumina support obtained as $\dot{Q} - \dot{Q}_{\mathrm{Cu}}(T)$ and assuming the $K_{\rm{al}} \propto T^{2.7}$ relation, we were able to determine $K_{\rm{al}}(T)$ by fitting the data above 0.25\,K to $K_{\rm{al}}/T^{2.7}=3.7(\dot{Q}-\dot{Q}_{\mathrm{Cu}}(T))/(T_{\mathrm{TO}}^{3.7}-T_{\mathrm{MC}}^{3.7})$ (see Fig.~\ref{fig-K}(c)).
The good fitting quality indicates the validity of the procedure.

The thermal conductivity deduced from the resultant $K_{\rm{al}}$ is one-half of the reported value~\cite{Locatelli1976}, presumably because of different grain sizes and sintering conditions of alumina. 
The lowest temperature recorded by the RuO$_2$ thermometer was 28\,mK when $T_{\mathrm{MC}} = 20$\,mK and $\dot{Q} = 0$.
This means that an ambient heat leak in our setup is only 5\,nW.
The lowest temperature will be further improved down to nearly 20\,mK if the copper weak link is replaced with a lead superconducting heat switch of appropriate design without degrading the thermal isolation around 1\,K~\cite{Montgomery1958, Reese1962}.

\section{Simulation of heat capacity measurements}
\label{simulation}

\begin{figure}[t]
\centering
\includegraphics[width=0.9\linewidth]{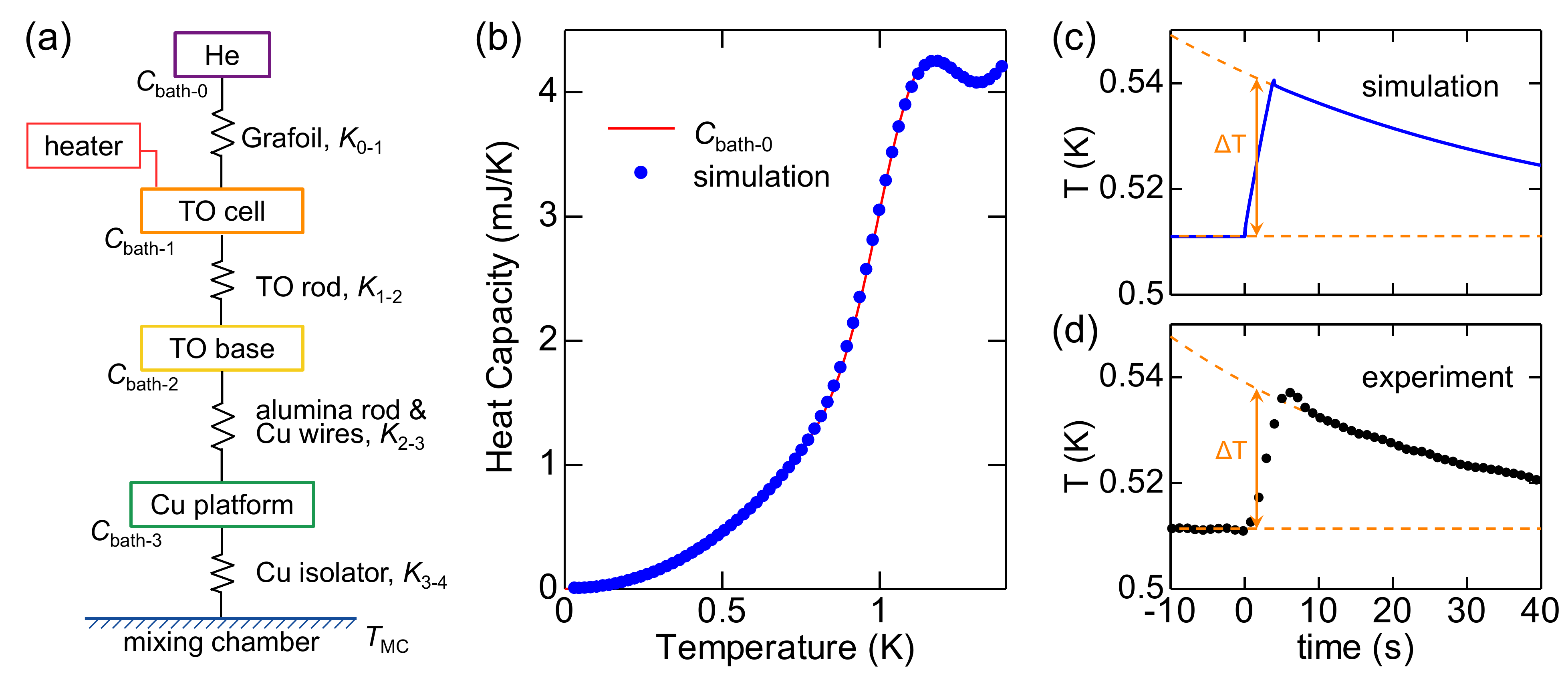}
\caption{(a) One-dimensional 3+1 bath model for thermal simulation. (b) The red line is the smoothed HC data for the $^4$He film of four layers (35\,nm$^{-2}$) taken from Ref.~\cite{Greywall1993}, which was input to $C_{\textrm{bath-0}}$ in the simulation after being normalized to the surface area of the present work. The blue dots are simulated HCs. (c) Example of the HC measurement simulation with the quasi-adiabatic heat pulse method. (d) Experimental HC data taken under the same conditions as those for (c). See the main text for further details.}
\label{fig-sim}
\end{figure}

Before carrying out actual HC measurement with the quasi-adiabatic heat pulse method, we have examined how it works from thermal simulations on a one-dimensional 3+1 bath model.
In this model (see Fig.~\ref{fig-sim}(a)), the following three thermal baths are connected in series; the TO cell consisting of the copper cap, Grafoil, and a BeCu main body, to which the heat pulse is injected ($C_{\textrm{bath-1}}$), a BeCu TO base plate ($C_{\textrm{bath-2}}$), and the copper platform ($C_{\textrm{bath-3}}$).
Finally, the third bath is connected to the mixing chamber held at a constant temperature ($T_{\mathrm{MC}}$).
Those baths are connected via three thermal conductances; the BeCu torsion rod ($K_{\textrm{1-2}}$), the alumina support and the copper weak  link ($K_{\textrm{2-3}}$), and the copper rod of the vibration isolator ($K_{\textrm{3-4}}$).
Known specific heats~\cite{Pobell2007,VanDerHoeven1966}, thermal conductivities~\cite{Pobell2007,Locatelli1976,Berman1955}, weights of materials used in the cell, and $K$ determined in the previous section are implemented into $C_{\textrm{bath-}i}$ and $K_{i\textrm{-}(i+1)}$, where $i = 1, 2, 3$.
Here, we approximated the specific heat of BeCu to that of copper. 
The model has an additional thermal bath of He film sample ($C_{\textrm{bath-0}}$), which is connected to $C_{\textrm{bath-1}}$ as seen in Fig.~\ref{fig-sim}(a) through the thermal conductance of Grafoil, taking account of its anisotropy~\cite{Dillon1985}.
The simulation is performed by solving numerically the following simultaneous equations of time evolution of the temperature of each bath-$i$:
\begin{equation}
C_{\textrm{bath-}i} \frac{d T_{i}}{d t}= \int_{T_{i}}^{T_{i-1}} K_{(i-1)\textrm{-} i} d T + \int_{T_{i}}^{T_{i+1}} K_{i\textrm{-} (i+1)} d T + \dot{Q}_{i}(t),
\end{equation}
where $\dot{Q}_i(t) = \dot{Q}$ during the heat pulse only for $i = 1$ otherwise $\dot{Q}_i = 0$ and also $K_{(i-1)\textrm{-} i} = 0$ for $i=0$.

In the simulation, $C_{\textrm{bath-0}}$ was set equal to smoothed HC data for the $^4$He film of four atomic layers (total areal density of 35\,nm$^{-2}$) on Grafoil measured by Greywall~\cite{Greywall1993} using a cell specialized for HC measurement (the solid line (red) in Fig.~\ref{fig-sim}(b)).
Fig.~\ref{fig-sim}(c) shows an example of such a simulation near 0.5\,K, where a time evolution of the temperature of the bath-1, $T_{\mathrm{TO}}(t)$, is simulated after applying a heat pulse of $\dot{Q} = 5.5$\,$\mu$W to it for $0 \leq t \leq 4$\,s.
Here one can see a little overshooting immediately after the heat pulse.
Thus, we evaluated the corresponding temperature increase $\Delta T$ by extrapolating the $T_{\mathrm{TO}}(t)$ recorded in a time span of $10 \leq t \leq 40$\,s back to the middle time of the heat pulse ($t = 2$\,s) as shown in the figure, and calculated the heat capacity as $C = Q/\Delta T$. 
The dots (blue) in Fig.~\ref{fig-sim}(b) are the simulation results obtained in such a way.
They perfectly agree with the input $C_{\textrm{bath-0}}$ ensuring the simulation procedure.
Note that the addendum HC calculated without the He bath has already been subtracted from the data points in Fig.~\ref{fig-sim}(b).

Fig.~\ref{fig-sim}(d) is a result of the HC experiments using our setup for a $^4$He sample of the same density and the same heat pulse as those for the simulation in (c).
Overall, the simulation reproduces the experiment well, indicating that the present heat bath model is appropriate.
The slightly dull response and a larger overshoot than the simulation are probably due to the existence of temperature gradients within the He sample and/or Grafoil in the actual apparatus.

\section{Test results of simultaneous TO and HC measurements}
\label{result}

\begin{figure}[b]
  \begin{center}
  \includegraphics[width=0.8\linewidth]{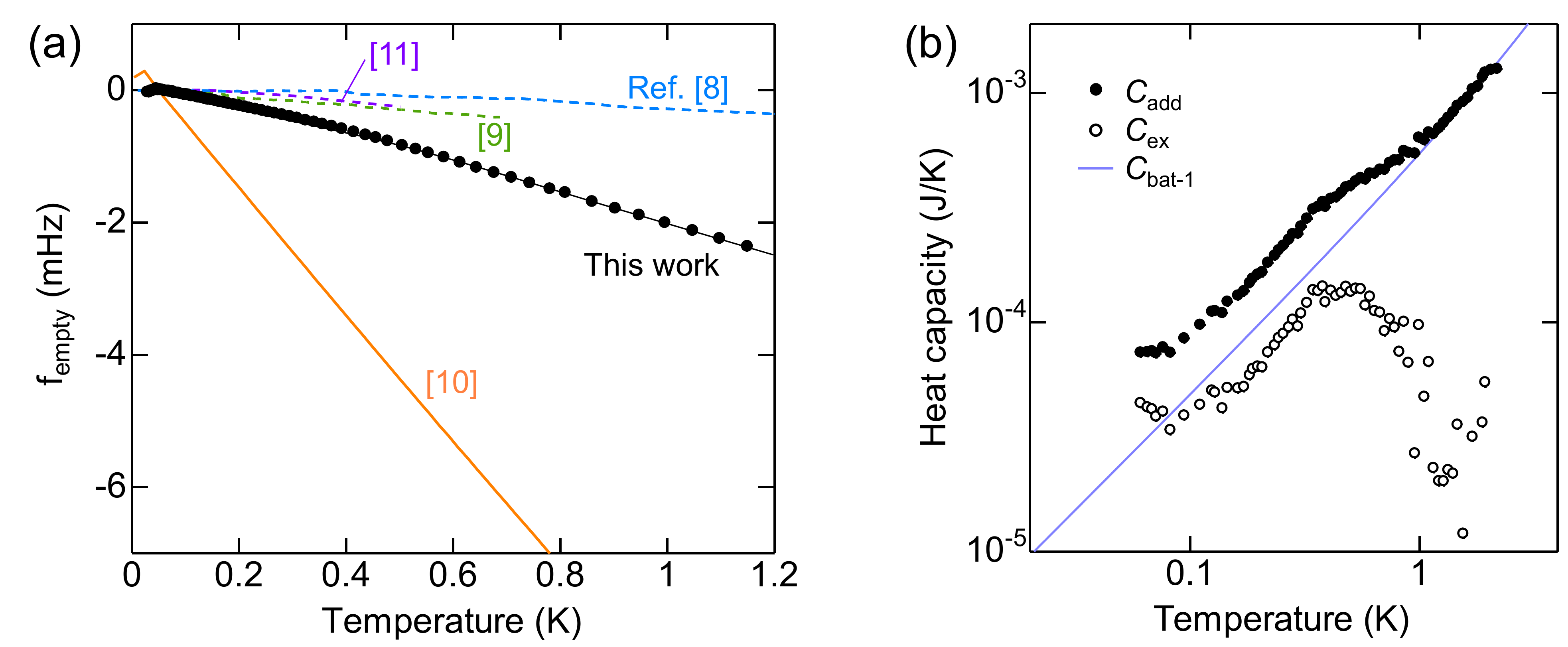}
  \caption{(a) Temperature dependence of the resonant frequency without helium sample ($f_{\mathrm{empty}}$) of our hybrid setup (dots) compared with those of the previous studies using BeCu TO rods~\cite{Crowell1996,Shibayama2009,Choi2021} (dashed line) and an AgCu TO rod~\cite{Nyeki2017} (solid line). Each frequency is shifted vertically so that $f_{\mathrm{empty}}=0$ at $T=50$\,mK. (b) Measured addendum HC ($C_{\mathrm{add}}$) of the hybrid cell (dots) and its expectation (solid line; $C_{\textrm{bath-1}}$). The open circles are the excess HC ($C_{\mathrm{ex}}= C_{\mathrm{add}} - C_{\textrm{bath-1}}$).}
  \label{fig-empty}
  \end{center}
\end{figure}

By measuring the resonance frequency of the empty cell ($f_{\textrm{empty}}$) and its $Q$ value, we confirmed that there was no measurable degradation in the TO performance after implementing the HC measurement capability into the original TO.
Then, we measured the temperature dependence of $f_{\textrm{empty}}$ below 1.2\,K as shown in Fig.~\ref{fig-empty}(a) (dots (black)).
Also shown here are those of the previous studies using cells specialized to TO measurement~\cite{Crowell1996,Shibayama2009,Nyeki2017,Choi2021}.
They all contain the Grafoil substrate.
It is important to keep the temperature variation of $f_{\textrm{empty}}$ as small as possible for reliable determination of the superfluid fraction.
Although our $f_{\textrm{empty}}$ has a little larger $T$-dependence among those using BeCu TO rods~\cite{Crowell1996,Shibayama2009,Choi2021}, it is a factor of four smaller than that using a coin silver TO rod~\cite{Nyeki2017}, indicating usability for our purpose.
We also found no mutual influences such as crosstalk during the heat pulse, meaning that genuinely simultaneous measurements are possible.

Next, we measured the addendum HC ($C_{\mathrm{add}}$) of this hybrid setup.
It is shown as the dots in Fig.~\ref{fig-empty}(b).
Above 1\,K, $C_{\mathrm{add}}$ is in good agreement with the expectation ($C_{\textrm{bath-1}}$) shown by the solid line (blue).
However, there is an excess HC ($C_{\textrm{ex}} \equiv C_{\mathrm{add}} - C_{\textrm{bath-1}}$) contribution below 1\,K with a rounded maximum near $T = 0.45$\,K (open circles).
The origin of the $C_{\textrm{ex}}$ is not known at present, but it could be related to tunneling degrees of freedom of deuterium impurities~\cite{Usami2021} in copper or BeCu.

%
\begin{figure}[t]
  \begin{center}
  \includegraphics[width=0.85\linewidth]{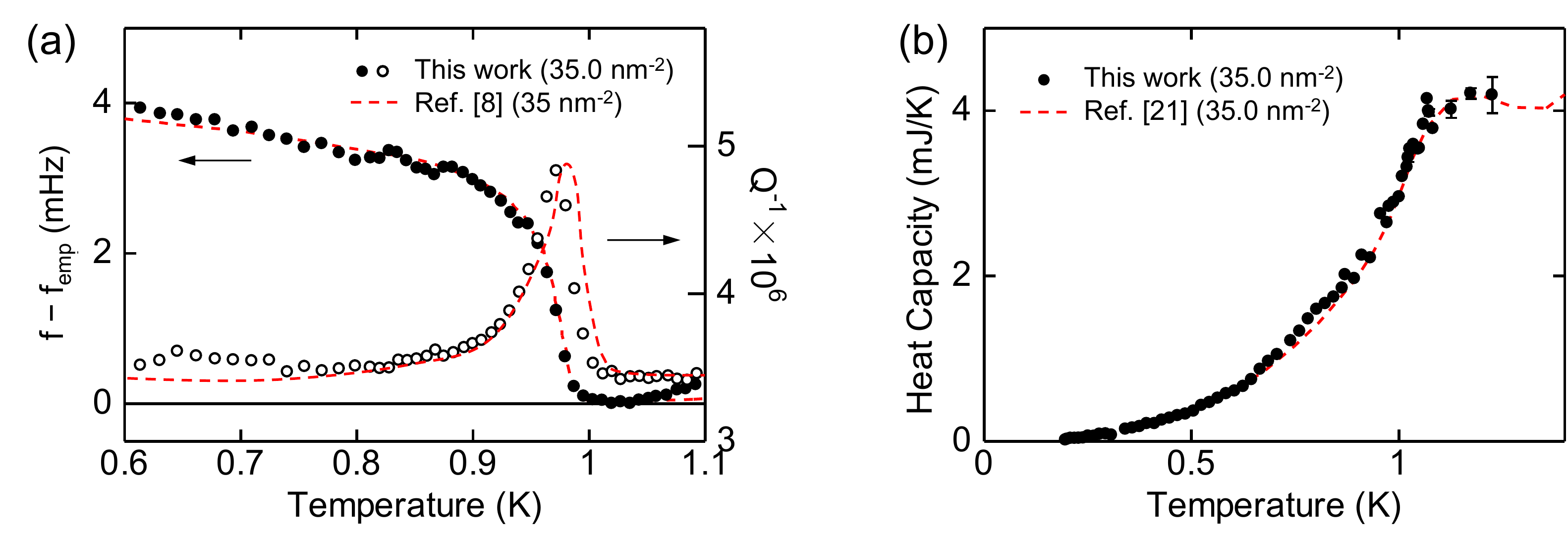}
  \caption{Results of the simultaneous (a) TO and (b) HC measurements for the $^4$He film of four layers (35\,nm$^{-2}$) on Grafoil. The dashed lines are the previous workers' data for similar density samples; Ref.~\cite{Crowell1996} in (a) and Ref.~\cite{Greywall1993} in (b). In (a), dots are frequency shifts from the value at $T = 1.02$\,K, and the open circles are the dissipation $Q^{-1}$.}
  \label{fig-KT}
  \end{center}
\end{figure}
%
 
Finally, we show a TO and HC data set taken on the same $^4$He film sample of 35\,nm$^{-2}$ in Figs.~\ref{fig-KT}(a)(b).
Here the background temperature dependence of $f$ and the addendum HC are already subtracted. 
This sample with a density close to fourth layer completion is considered to consist of two solid layers and two liquid layers.
The liquid overlayers are believed to undergo a superfluid state through the Kosterlitz-Thouless (KT) transition at $T \approx 1$\,K~\cite{Crowell1996,Greywall1993}.
Our new TO data clearly show the KT transition at $T_{\textrm{KT}} = 0.98$\,K where a dissipation ($Q^{-1}$) peak and an $f$ jump are observed (Fig.~\ref{fig-KT}(a)).
On one hand, the HC shows an exponentially rapid decrease below a broad peak at a sightly higher temperature than $T_{\textrm{KT}}$ as seen in Fig.~\ref{fig-KT}(b).
This is consistent with the theoretical expectation from the vortex unbinding mechanism of the KT theory~\cite{Berker1979}.
It is important that the original KT idea is verified both from the superfluid response and thermodynamic measurements on the same sample.

The dashed lines (red) in Figs.~\ref{fig-KT}(a)(b) are the previous workers' results~\cite{Crowell1996,Greywall1993} on samples with nearly the same densities as ours. 
The TO data by Crowell-Reppy~\cite{Crowell1996} in (a) and the HC data by Greywall~\cite{Greywall1993} in (b) are normalized by the apparent frequency jump at $T_{\textrm{KT}}$ and the surface area of Grafoil, respectively, so as to be consistent with our data.
Clearly, our new data set agrees very well with a set of these de facto standard previous experiments, which were performed independently by different groups.

\section{Conclusions}
We developed a simple method to make possible heat capacity (HC) measurement using an existing torsional oscillator (TO).
The method involves insertion of a rigid alumina support in between the TO cell and the vibration isolator in order to thermally isolate the cell keeping the rigidity of the whole equipment.
For our specific purpose, a copper-wire weak link is also put in parallel with the alumina support so that the TO and HC measurements can be extended down to 30\,mK while the reliable high-$T$ HC measurement around 1\,K can also be made.
We described how to design the apparatus and to test the performances both by simulation and experiment in detail. 
The capability of simultaneous TO and HC measurements with this hybrid apparatus was nicely demonstrated by measuring superfluid responses and a broad heat capacity anomaly in multilayer $^4$He film adsorbed on graphite, which are consistent with the KT theory for a pure two-dimensional superfluid transition.
This is a practical and useful technique when simultaneous TO and HC measurements on exactly the same superfluid helium sample are required.

\bmhead{Acknowledgments}

\ 
This work was financially supported by Japan Society for the Promotion of Science (JSPS) KAKENHI Grant Number JP18H01170. 
H.F. thanks Eunseong Kim for suggesting the importance of simultaneous TO\&HC measurements for studies of superfluid $^4$He adlayers on graphite. 
J.U. appreciates Satoshi Murakawa for useful discussions, and JSPS for supports through Program for Leading Graduate Schools (MERIT) and Grant-in-Aid for JSPS Fellows JP20J12304.





\bibliography{Ref}


\end{document}